# Exploring mechanical and thermal properties of high-entropy ceramics via general machine learning potentials

Yiwen Liu, Hong Meng, Zijie Zhu, Hulei Yu[*], Lei Zhuang, Yanhui Chu[*]

School of Materials Science and Engineering, South China University of Technology,

Guangzhou, 510641, China

[*] Corresponding author.
*E-mail address*: huleiyu@scut.edu.cn (H. Yu); chuyh@scut.edu.cn (Y. Chu)

# Abstract

The mechanical and thermal performance of high-entropy ceramics are critical to their use in extreme conditions. However, the vast composition space of high-entropy ceramic significantly hinders their development with desired mechanical and thermal properties. Herein, taking high-entropy carbides (HECs) as the model, we show the efficiency and effectiveness of exploring the mechanical and thermal properties via machine-learning-potential-based molecular dynamics (MD). Specifically, a general neuroevolution potential (NEP) with broad compositional applicability for HECs of ten transition metal elements from group IIIB-VIB is efficiently constructed from the small dataset comprising unary and binary carbides with an equal amount of ergodic chemical compositions. Based on this well-established NEP, MD simulations on mechanical and thermal properties of different HECs have shown good agreement with the results of first-principles calculations and experimental measurements, validating the accuracy, generalization, and reliability of using the developed general NEP in investigating mechanical and thermal performance of HECs. Our work provides an efficient solution to accelerate the search for high-entropy ceramics with desirable mechanical and thermal properties.

# 1. Introduction

High-entropy ceramics, a class of emergent multicomponent materials with four or more principal elements, have attracted substantial attention in the ceramic community since they were first proposed in 2015 [1-5]. Owing to their compositional complexity and unique microstructures, high-entropy ceramics have been shown to possess a range of superior properties exceeding those of their individual component ceramics, such as better thermal stability [1,2,6,7], enhanced mechanical properties [3,4,8,9], exceptional oxidation and ablation resistance [10-13], higher catalytic activity [14,15], improved thermoelectricity [16,17], and remarkable superionic conductivity [18]. These distinctive characteristics have made them great potential for various structural applications [1-4,6-13], such as hypersonic thermal protection systems and high-speed cutting tools, as well as functional applications [14-18], like catalysts, thermoelectrics, and electrodes.

The investigations of the mechanical and thermal properties are essential for high-entropy ceramics' use as structural materials, especially in extreme circumstances. To date, various exceptional mechanical and thermal properties, like high hardness and modulus [3,4,9], enhanced flexural strength [19,20], improved high-temperature creep resistance [21], lowered thermal conductivity [6,7,16], and matched coefficients of thermal expansion (CTEs) with silicon-based substrates [22], have been reported in high-entropy ceramics. Nevertheless, research on the mechanical and thermal properties of high-entropy ceramics is still in the early stages. Due to the large composition space, the traditional experimental approaches to searching for high-entropy ceramics with desirable mechanical and thermal characteristics are both time-consuming and costly, failing to meet the growing demand from future structural application domains. Hence, it is imperative to develop an efficient method to comprehensively explore the mechanical and thermal properties of high-entropy ceramics. To this end, molecular dynamics (MD) simulations based on machine learning potentials (MLPs), as a frontier data-driven approach, have emerged as an effective and powerful solution [23-25]. Although studies of MLPs in predicting material properties have been carried out in high-entropy alloys (HEAs) [26-28] and ceramic systems, like $TiO_2$ [29], $Ti_3O_5$ [30], and $TiB_2$ [31], the applications of MLPs in high-entropy ceramics are seldom reported, with only two attempts proposed by Dai et al. on constructing MLPs for $(Ti_{0.2}Zr_{0.2}Hf_{0.2}Nb_{0.2}Ta_{0.2})B_2$ [32] and $(Zr_{0.2}Hf_{0.2}Ti_{0.2}Nb_{0.2}Ta_{0.2})C$ [33]. The dominant constraint lies in the poor transferability of the trained MLP with the conventional construction strategy, where the MLP is trained from a fixed-composition dataset, greatly hindering the comprehensive predictions of mechanical and thermal properties in different materials. Recent studies have demonstrated improved transferability in MLPs trained with varied compositions in HEAs, promoting MD simulations on their plasticity and primary radiation damage performance [34]. However, there is still a lack of general MLPs with broad compositional applicability to enable the effective and economical exploitation of high-entropy ceramics with desired mechanical and thermal properties.

High-entropy carbides (HECs) are promising ultrahigh-temperature ceramics due to their potential tailorable and remarkable mechanical and thermal performance [3,4,8,9]. In this work, taking the HEC system as an example, we systematically explore the mechanical and thermal properties of HECs. To be specific, a general neuroevolution potential (NEP) with high accuracy and transferability based on unary (1HEC; the carbides with *n* types of transition metal (TM) elements are notated as *n*HECs) and binary (2HEC) carbide training data with up to ten TM elements of group IIIB, IVB, VB, and VIB (see **Fig. 1**) is efficiently built, enabling accurate MD simulations on properties HECs. Various effects, including the computational cost and modeling complexity of density functional theory (DFT) computations as well as accuracy and transferability of NEPs trained from different datasets, on the construction of NEPs are discussed, and a combination of 1HEC and 2HEC (1+2HEC) configurations with an equal amount of all possible chemical compositions is identified as the optimal training dataset choice. The high transferability and accuracy of the established NEP are then assessed and verified by considering its applicability to HECs. Moreover, the accuracy, generalization, and reliability of MD simulations with the trained NEP on predicting the mechanical and thermal properties of HECs are validated by comparing the results of first-principles calculations and experimental measurements. Our established NEP can efficiently accelerate the discovery of HECs with desired mechanical and thermal properties via MD simulations.

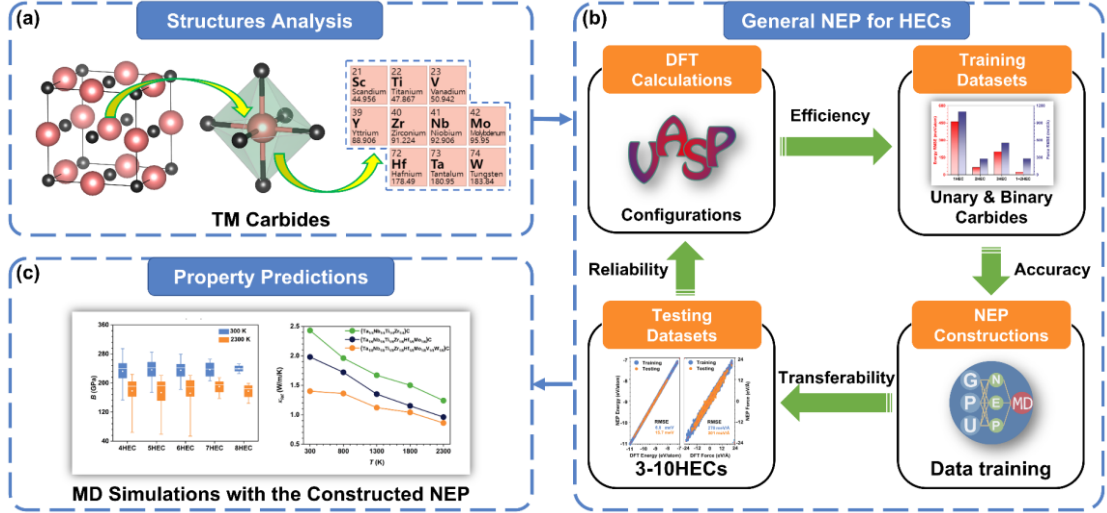

**Figure 1** Schematic of the strategy to construct general NEP for HECs. (a) Crystal structure and the polyhedral of TM carbides with ten TM elements. TM and carbon atoms are colored in pink and black, respectively. (b) NEP constructions for HECs. (c) MD simulations on mechanical and thermal properties of HECs based on the trained NEP.

## 2. Experimental and computational methods

**2.1 DFT calculations**

All DFT calculations were implemented in the Vienna *ab initio* simulation package (VASP) with the projector augmented wave (PAW) [35,36]. The Perdew-Burke-Ernzerhof (PBE) under the generalized gradient approximation (GGA) was applied to approximate the electronic exchange and correlation function [37]. Supercells of 1-10HECs were built based on the ZrC conventional cell, and the special quasi-random structure (SQS) approach was performed in the Alloy Theoretic Automated Toolkit (ATAT) [38,39] for 2-10HECs. Detailed supercell sizes are summarized in **Table S1**. All the structures were fully relaxed initially to obtain the lattice constants. Additionally, elastic tensors from DFT were calculated using the energy-strain approach [40]. Voigt-Reuss-Hill method was also used to evaluate the bulk modulus (*B*) and the shear modulus (*G*) of HECs [41].

To build training and testing datasets, the *ab initio* molecular dynamics (AIMD) calculations [42] with only the Γ point in the Brillouin zone considered were performed. A low plane-wave energy cutoff of 300 eV was used to accelerate the collection of enough different structural configurations. A time step of 3 fs and a total simulation time of 6 ps were applied under the NPT ensemble. The simulated temperatures were set to 1000, 2000, 3000, and 4000 K, respectively. For high-accurate single-point calculations of configurations, the plane-wave cutoff energy was set to 450 eV, and the Brillouin zone was sampled employing the Γ-centered method with a separation of approximately 0.5 Å$^{-1}$ [43]. The energy convergence criterion of the electronic self-consistency cycle was set to $10^{-5}$ eV.

For the AIMD simulations of the tensile deformation along different directions, SQS supercell models with x || [$\bar{1}$10], y || [$\bar{1}$10], z || [001] and x || [111], y || [$\bar{1}$2$\bar{1}$], z || [$\bar{1}$0$\bar{1}$], respectively, were constructed (96 atoms for [100], 192 atoms for [110], and 144 atoms for [111]). To ensure the accuracy and efficiency of the simulations, a time step of 1 fs and a cutoff energy of 400 eV were applied. Equilibration of all models at 300 K was first obtained for 1 ps relaxation under the NPT ensemble, followed by 1ps NVT simulations, and then the equilibrated models were elongated along the corresponding direction with a 2% increment in strain. All the tensile supercells were then equilibrated under the NPT ensemble with the fixed directional elongation for 1 ps at each

deformation step. The averaged stress tensors were collected for the final 0.1 ps.

## 2.2 NEP constructions

Small-scale 1HEC, 2HEC, 3HEC, and 1+2HEC training datasets were first developed with each comprising 1000 configurations from AIMD calculations. The distribution of different TM elements in the datasets was ensured with the highest degree of equality. Different numbers of 1HEC configurations were considered in the small-scale 1+2HEC training datasets with ratios of 18.18% (equal numbers for all chemical compositions of 1HECs and 2HECs), 40%, 60%, and 80%. Based on these training datasets, state-of-the-art NEPs were then trained within the Graphics Processing Units Molecular Dynamics (GPUMD) code [44]. The detailed training hyperparameters are listed in **Table S2**. In addition, a corresponding testing dataset with 1-10HECs was constructed by a total of 300 randomly selected configurations (30 configurations for each $n$HEC) to evaluate the performance of the trained NEPs from these small-scale training datasets.

A large 1+2HEC training dataset containing 55 kinds of chemical compositions and 5500 configurations (100 configurations for each composition collected from AIMD) was built to train the accurate NEP for HECs. Meanwhile, 1500 configurations of randomly picked 71 kinds of HECs (ten kinds of chemical compositions each from 3-9HECs and the unique composition from 10HEC) were collected to build the 3-10HEC testing dataset with the best assurance of equally distributed TM elements. 20 configurations in each of the 3-9HEC compositions and 100 configurations in the 10HEC composition were selected from AIMD at the abovementioned four temperatures for static calculations.

## 2.3 Molecular dynamics simulations

The Large Scale Atomic/Molecular Massively Parallel Simulator (LAMMPS) package was utilized to perform MD simulations interfaced with the NEP code [45]. All the simulated models were built by randomizing the TM elements in a 20×20×20 FCC supercell (a total of 64000 atoms). A time step of 1 fs was maintained throughout the simulations. Periodic boundary conditions were used in all directions. All models were first equilibrated using the NVT ensemble for a period of 10 ps, and then relaxed using the NPT ensemble for a period of 10 ps, and finally equilibrated using the NVT ensemble for a period of 10 ps at 300 K. Temperature-dependent elastic tensors were computed at 300 K and 2300 K, respectively, by taking Born term into account within 10 ps. The magnitude of strain was set to $1.0E^{-6}$. For a better comparison, elastic tensors of DFT SQS models using energy-strain methods were also evaluated with the NVT ensemble for 10 ps. The MD simulation setup of tensile deformation was performed at an engineering strain rate of 0.01 $ps^{-1}$ for 20 ps. To consider the size effects, both the tensile deformations of the previous-mentioned SQS models and nanoscale models (64000 atoms for [100], 128000 atoms for [110], and 162000 atoms for [111], respectively) were explored with the above-mentioned model equilibration. The lattice thermal conductivity ($\kappa_{lat}$) was predicted by using the GPUMD package through the homogeneous nonequilibrium molecular dynamics (HNEMD) method [44] with a timestep of 1 fs and periodic boundary conditions. All models were equilibrated for 20 ps using the NPT ensemble with a target temperature of 300 K before collecting the following 2 ns heat current data in the NVT ensemble. During HNEMD simulations, the driving force was chosen as $F_e$ = 0.0001 Å$^{-1}$, and five independent simulations for each sample were performed to calculate the average thermal conductivity. A total correlation step of 200, a maximum angular frequency of 400 THz, and a sample interval of 2 were used for spectral thermal conductivity ($\kappa(\omega)$) calculations. In addition, the mass fluctuations were computed as follows:

$$\text{Mass fluctuations} = \sum_i \left| \frac{m_i - \bar{m}}{\bar{m}} \right| \tag{1}$$

where $m_i$ is the atomic mass of the $i$-th element, and $\bar{m}$ is the average mass of all elements. The atomic strain tensor ($\varepsilon_{ab}^i$) was measured using the Green-Lagrangian strain tensor, and the volumetric strain and shear strain were defined as follows [46]:

$$\text{Volumetric strain} = \frac{\varepsilon_{xx}^i + \varepsilon_{yy}^i + \varepsilon_{zz}^i}{3} \tag{2}$$

$$\text{Shear strain} = \sqrt{{\varepsilon_{xy}^i}^2 + {\varepsilon_{xz}^i}^2 + {\varepsilon_{yz}^i}^2 + \frac{(\varepsilon_{xx}^i - \varepsilon_{yy}^i)^2 + (\varepsilon_{xx}^i - \varepsilon_{zz}^i)^2 + (\varepsilon_{yy}^i - \varepsilon_{zz}^i)^2}{6}} \tag{3}$$

Moreover, equilibrium lattice constants from 300 K to 2300 K with an interval of 500 K were obtained, and the corresponding quadratic functions were fitted to evaluate CTEs at different temperatures. The melting point was simulated by first equilibrating at 2000 K for 100 ps using the NVT ensemble before being elevated to 4000 K with a heating rate of 5 K/ps using the NPT ensemble. The cohesive energy was computed by minimizing the potential energy of the HEC cells:

$$\text{Cohesive energy} = \sum_i E_i - E_{HEC} \tag{4}$$

where $E_i$ and $E_{HEC}$ are the potential energy of the $i$-th atom and the minimized HEC cell. The Pearson correlation between features and was evaluated to reveal the correlation between different properties [47]:

$$\text{Pearson correlation} = \frac{\sum(x_i - \bar{x})(y_i - \bar{y})}{\sqrt{\sum(x_i - \bar{x})^2 \sum(y_i - \bar{y})^2}} \qquad (5)$$

where $x_i$ and $y_i$ are the $i$-th values of two different input features, respectively. $\bar{x}$ and $\bar{y}$ are the expectations of the two different input features, respectively.

**2.4 Experimental methods**

The high-quality $(Ta_{1/4}Nb_{1/4}Ti_{1/4}Zr_{1/4})C$, $(Ta_{1/6}Nb_{1/6}Ti_{1/6}Zr_{1/6}Hf_{1/6}Mo_{1/6})C$, $(Ta_{1/8}Nb_{1/8}Ti_{1/8}Zr_{1/8}Hf_{1/8}Mo_{1/8}V_{1/8}W_{1/8})C$ samples were fabricated via a two-step strategy involving ultrafast high-temperature synthesis and hot-press sintering techniques. The detailed synthesis process can be found in our previous work [48]. The CTEs of the samples (13.5 mm × 3 mm × 3 mm) were measured over the temperature range of 373-998 K at a heating rate of 5 K/min in a nitrogen atmosphere by a high-temperature dilatometer (Netzsch DIL 402C, NETZSCH, Selb, Germany). The referenced temperature for calculating CTEs was 298 K. The thermal diffusivity ($h$) was measured at room temperature under an argon flow by a laser-flash apparatus (Netzsch LFA 427, Netzsch, Selb, Germany). The total thermal conductivity ($\kappa_{tot}$) of the samples was evaluated according to the following equation [49]:

$$\kappa_{tot} = h \cdot \rho \cdot C_p \qquad (6)$$

where $C_p$ is the specific heat capacity estimated based on the Dulong-Petit law and $\rho$ is the density of the samples measured by the Archimedes drainage method. As the densities of all samples were almost the same as theoretical densities, no correction on $\kappa_{tot}$ was considered. In addition, the electrical conductivity ($\sigma$) of samples at room temperature was measured on a commercial apparatus (CTA-3, Cryoall, Beijing, China) under a helium atmosphere, and the corresponding electronic contribution of thermal conductivity ($\kappa_{ele}$) was computed according to Wiedemann-Franz law [50]:

$$\kappa_{ele} = L \cdot \sigma \cdot T \tag{7}$$

where $L = 2.44 \times 10^{-8}$ W Ω/K² is the Lorenz number, and $T$ is the absolute temperature. The lattice thermal conductivity ($\kappa_{lat} = \kappa_{tot} - \kappa_{ele}$) was thereby obtained.

## 3. Results and discussion

The strategy of building general MLPs for HECs was first discussed. Unlike HEAs, where the metal principal elements are closely packed together, the metal elements in high-entropy ceramics are always separated by non-metal elements. As can be seen from Fig. 1(a), the first-nearest-neighbor atoms of TM atoms in carbides are all carbon atoms, forming an octahedron with TM-C bonding. When it comes to HEC systems (see **Fig. 2**(a)), such a simple atomic environment can also be perpetuated. Though there exists tiny local lattice distortion in specific atomic pair distance, the overall radial pair distribution function presented in Fig. 2(b) still confirms the similarity of the atomic environment for both TM carbides and HECs, which is beneficial to the construction of MLP descriptors for different HECs [23]. This indicates the possible applicability of MLPs between training datasets of carbides and their solid solutions. Based on these findings, it is anticipated that transferable MLPs applicable to various *n*HECs can be efficiently yielded from the training datasets comprising configurations of limited types of *n*HECs.

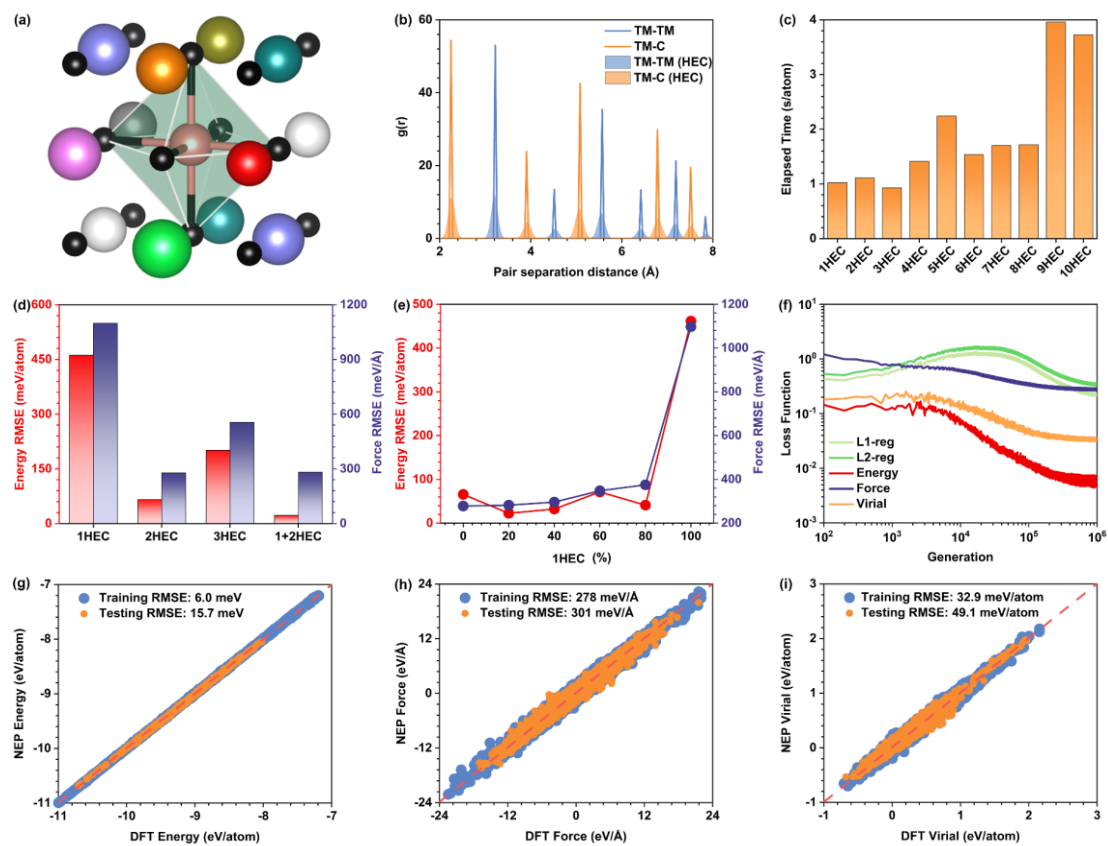

**Figure 2** Construction and performance of NEP for HECs. (a) Local crystal structure and polyhedral of HECs. Black atoms represent the carbon element, and other colored atoms represent different TM elements. (b) Radial pair distribution function of TM carbides and HECs with the same lattice constants. (c) Average elapsed time of $n$HECs in DFT calculations. (d) Performance of NEPs with different training datasets (small-scale) to 1-10HEC testing datasets. (e) Performance of NEPs with different 1+2HEC training datasets to 1-10HEC testing datasets. (f) Evolution of various loss functions with respect to generations for the constructed NEP from the 1+2HEC training dataset (large-scale). (g) Energy, (h) force, and (i) virial from NEP and DFT calculations for the 1+2HEC training dataset (large-scale) and the 3-10HEC testing dataset.

To verify this hypothesis, the cost of building datasets and the performance of trained NEPs from different datasets were first investigated. In HECs, the number of TM elements has a significant influence on geometric modeling, resulting in different efficiencies in DFT calculations for each composition. As shown in Fig. 2(c), it is found that the elapsed times show an overall increasing trend from 1HECs to 10HECs, indicating an unfavorable efficiency in building training datasets with more TM elements in HECs. Such computational inefficiency can also be explained by the relatively lower symmetry in HECs with more TM elements. As 1-3HEC configurations have the highest computational speeds, NEPs were first trained based on small-scale training datasets of 1HEC, 2HEC, and 3HEC, respectively, and their performances are presented in **Fig. S1-S3** and Fig. 2(d). While the trained NEPs all possess high training accuracy with comparable root mean square errors (RMSE) in energy (3.7-4.3 meV/atom), force (245-273 meV/Å), and virial (28.2-34.6 meV/atom), their transferability to the 1-10HEC testing dataset is quite different. Among the three trained NEPs, the one trained from the 1HEC training dataset performed the worst, showing low transferability in 1-10HEC systems. In contrast, NEP from the 2HEC training dataset has better testing accuracy with significantly lower RMSEs of energy (65.6 meV/atom) and force (278 meV/Å), which indicates the enhanced transferability for *n*HEC systems and verifies our hypothesis. Furthermore, it is notable from Fig. 2(e) and **Fig. S4-S7** that NEPs trained from the mixed 1+2HEC training datasets demonstrate better accuracy and transferability than both NEPs from 1HEC and 2HEC training datasets, possessing much lower RMSEs of energy with a reduction of up to 65.7% besides the comparable RMSE of force. The best ratio of 1HEC: (1HEC+2HEC) is identified to be 18.18%, where equal numbers for all chemical compositions of 1HECs and 2HECs were used in training. In addition, ten TM elements can yield a total of 1023 equimolar compositions in 1-10HECs. For the 1+2HEC dataset, merely ten kinds of 1HECs and 45 kinds of 2HECs need to be taken into consideration, significantly simplifying the modeling processes for AIMD calculations. Therefore, the 1+2HEC dataset with an equal amount of all possible chemical compositions is determined to be the most efficient training dataset.

The accuracy of the trained NEP can be further improved by increasing the volume of the training dataset [25]. Here, a larger 1+2HEC training dataset, configurations increasing from 1000 to 5500, was built to construct the NEP with high accuracy. As shown in Fig. 2(f), the L1 and L2 regularization loss functions show an increasing-then-decreasing trend over one million generations, indicating the effectiveness of the regularization. The loss of energy, force, and virial also converged, suggesting the well-trained NEP. In addition, the increase in the volume of the 1+2HEC training dataset has no significant influence on their training accuracy (see Fig. 2(g-i)). To examine the transferability of the trained NEP, a testing dataset based on 1500 kinds of 3-10HECs was built. It is notable that the enlarged volume of the training dataset can greatly improve the testing accuracy, achieving low testing RMSEs of 15.7 meV, 301 meV/Å, and 49.1 meV/atom in energy, force, and virial, respectively. Although the accuracy of the trained NEP is promising to be optimized further by introducing more training data (tens of thousands for most previous MLP works [29,33]), the present testing accuracies have already been sufficiently high for physical and chemical property explorations compared to previously reported MLPs [30,32-34], demonstrating the remarkable efficiency, transferability, and accuracy of our established NEP for HECs. Moreover, only a minor fluctuation in testing accuracies can be observed in individual testing accuracies of $n$HEC ($n$ = 3-10) testing datasets from **Fig. S8**, indicating the generalization performance of our trained NEP to different HEC systems.

With the established NEP, MD simulations can efficiently explore the performance of the mechanical and thermal properties of HECs. A total of 200 compositions of different 4-8HECs were randomly selected. As displayed in **Fig. 3**(a), the predicted lattice parameters ($a$) of 4-8HECs from MD simulations at 300 K are well consistent with DFT results and reported experiments [7,49-54], indicating the reliability of our trained NEP. From Fig. 3(b-f), it can be found that the three elastic constants ($c_{11}$, $c_{12}$, and $c_{44}$) of HECs and the corresponding two modulus ($B$ and $G$) based on these constants predicted by NEP without temperature and size effects exhibit remarkable alignment to those of DFT computations with the negligible value difference (see RMSE in Fig. 3(g)), showing the accuracy of our fitted NEP. On the contrary, large discrepancies can be observed in Fig. 3(b-g) between temperature-dependent MD results (300 K) and DFT calculations. Such large disparities demonstrate the importance of the size effects [31] (more disordered model) and temperature effects on the evaluations of the elastic properties of HECs, thereby, implying the necessity of establishing accurate MLPs for HEC property simulations. Moreover, the elastic properties of different $n$HECs ($n$ = 4-8) at both high temperature and low temperature were also explored. As shown in Fig. 3 (h,i), it can be seen that the differences in the average $B$ and $G$ are tiny for 4-8HECs, and HECs with fewer TM elements tend to show larger value ranges at both low and high temperatures. Furthermore, the increase in temperature is determined to be detrimental to $B$ and $G$ for HECs with an obvious overall decrease from 300 K to 2300 K. As a result, conclusions can be drawn that the established NEP has high accuracy and good generalization performance in MD simulations of HECs.

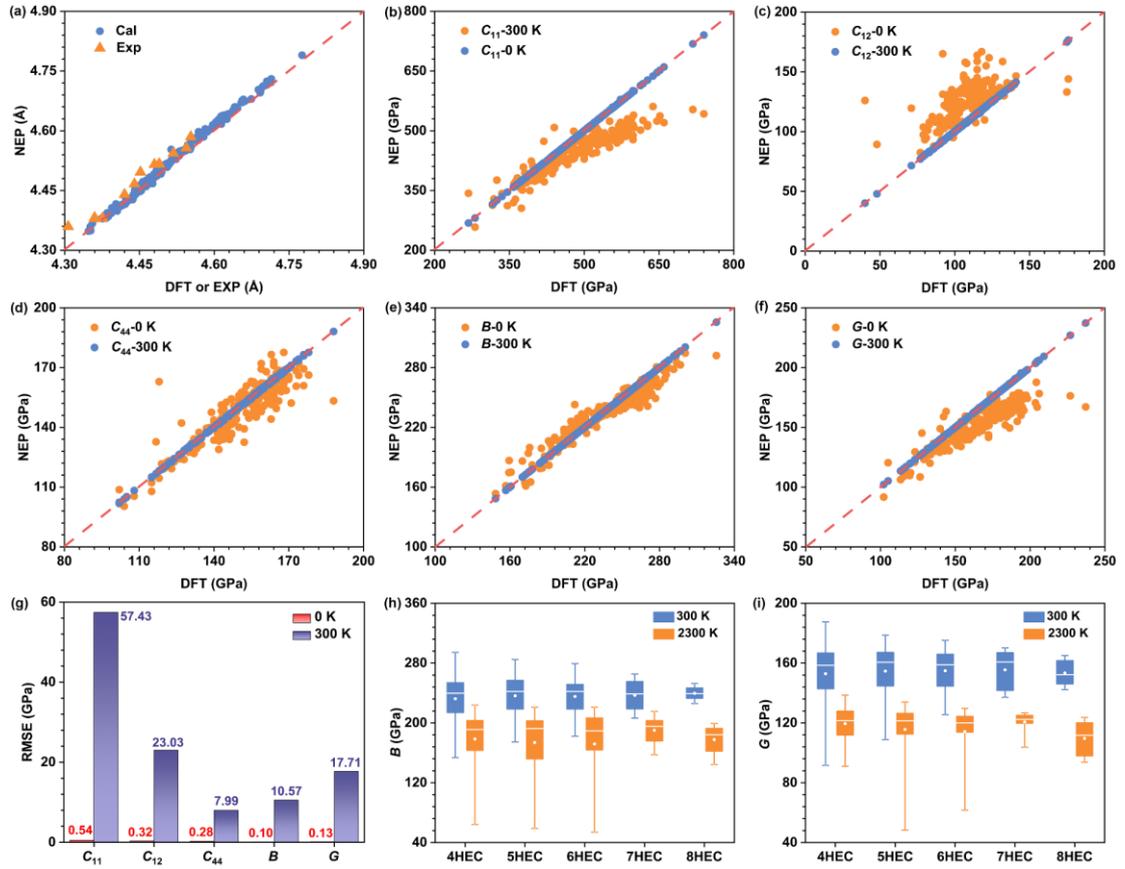

**Figure 3** Comparison of structural and elastic properties of 4-8 HECs between NEP and DFT calculations. (a) Lattice constants. Experimental data are from Ref. [7,49-54]. (b-d) Elastic tensors $c_{11}$, $c_{12}$, and $c_{44}$. (e) $B$. (f) $G$. (g) RMSE for the above five quantities from NEP calculations with and without considering temperature effects. (h) $B$ of 4-8HECs at 300 K and 2300 K. (i) $G$ of 4-8HECs at 300 K and 2300 K.

Besides the structural and elastic properties, the applicability of our established NEP on the tensile strengths of 4HEC was also investigated. Taking the common 4HEC $(Ta_{1/4}Nb_{1/4}Ti_{1/4}Zr_{1/4})C$ [48] as an example, **Fig. 4** presents the simulated tensile-testing results at 300 K. As depicted in Fig. 4(a-c), the ultimate tensile strengths of $(Ta_{1/4}Nb_{1/4}Ti_{1/4}Zr_{1/4})C$ from MD simulations, which are the global stress maximum during the tensile tests [31], are approximately 25.15 GPa, 37.00 GPa, and 35.41 GPa, for the [100], [110], and [111] directions, respectively, which is in agreement with the ones from AIMD values (22.97 GPa, 38.34 GPa, and 34.55 GPa, respectively). Notably, the slight discrepancies in results at high strains may be attributed to the underfitting of the datasets without fractured configurations. Meanwhile, the tensile toughness was also evaluated from the area underneath stress-strain curves [31]. The NEP results are 1.58 GPa, 3.06 GPa, and 2.87 GPa for the [100], [110], and [111] directions, respectively, which are also comparable to those of AIMD results (1.43 GPa, 3.08 GPa, 2.88 GPa, respectively). In addition, the predicted tensile strength and toughness from nanoscale models vary a lot, indicating the great influence of size effects similar to elastic properties, especially approaching the maximum points with large strains. Moreover, the fracture surfaces formed during different deformations from MD simulations (see Fig. 4(d-f)) are also aligned with the results of AIMD (see Fig. 4(g-i)), which are all approximately perpendicular to loading directions. It is notable that tensile loading along [100] will yield a void first in the fracture process, and the [110] tensile loading opens a zigzag fracture surface. Such exceptional quantitative agreement of stress-strain properties between MD and AIMD simulations indicates the reliability of our trained NEP.

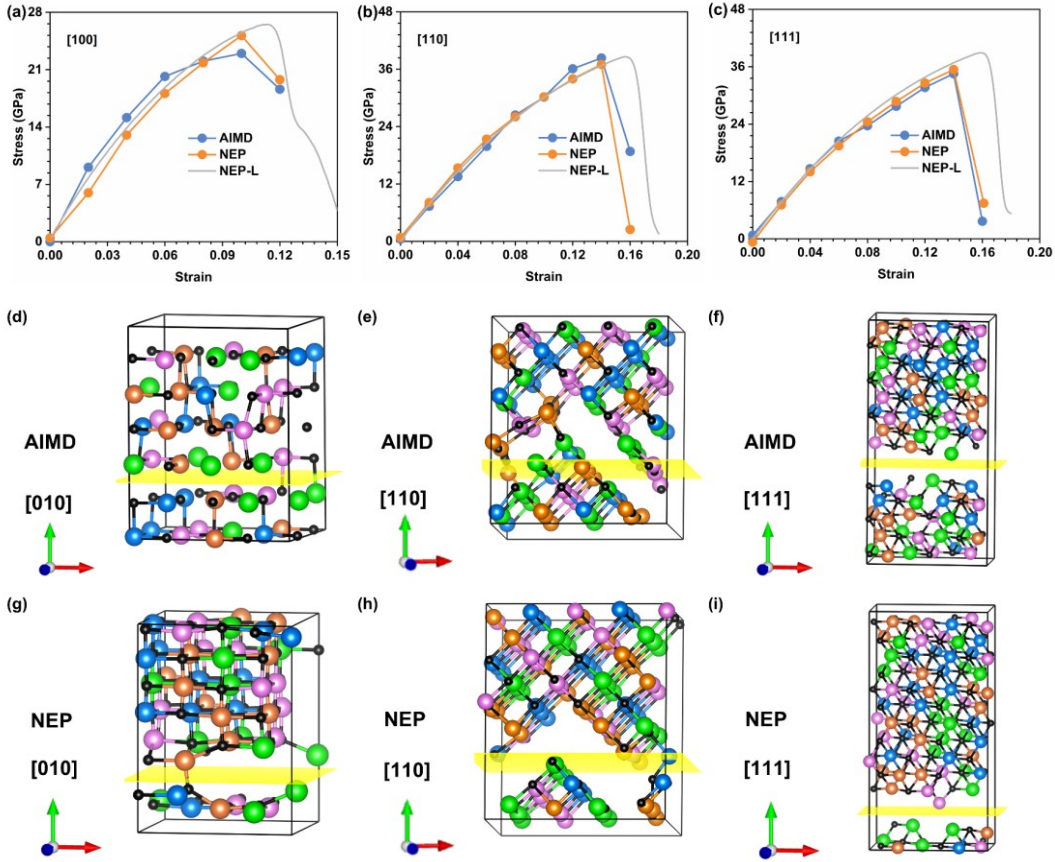

**Figure 4** Tensile deformation for $(Ta_{1/4}Nb_{1/4}Ti_{1/4}Zr_{1/4})C$ with NEP and AIMD calculations. (a-c) Stress-strain curves of SQS models and nanoscale models (NEP-L) at 300 K. (d-f) Snapshots of fracture surfaces on (100), (110), and (111) planes from AIMD (0.12, 0.16, and 0.16 strain, respectively). (g-i) Snapshots of fracture surfaces on (100), (110), and (111) planes from NEP (0.12, 0.16, and 0.16 strain, respectively).

To validate the reliability of our trained NEP in studying the thermal properties of HECs, the predictions on thermal conductivities of three selected HECs at 300 K were first carried out. It should be noted that the electronic contribution ($\kappa_{ele}$) has a significant influence on $\kappa_{tot}$ of HECs (see **Table S3**) due to the gapless band structure of carbides [55], while the one obtained from our MD simulations is merely the lattice part ($\kappa_{tot}$). **Fig. S9** shows the running $\kappa_{lat}$ as a function of correlation time, where well-converged averaged $\kappa_{lat}$ can be obtained in the range between 1.0 and 2.0 ns, indicating the reliability of our HNEMD simulations. Meanwhile, it is found that there exists a slight decrease from (Ta$_{1/4}$Nb$_{1/4}$Ti$_{1/4}$Zr$_{1/4}$)C to (Ta$_{1/6}$Nb$_{1/6}$Ti$_{1/6}$Zr$_{1/6}$Hf$_{1/6}$Mo$_{1/6}$)C and then to (Ta$_{1/8}$Nb$_{1/8}$Ti$_{1/8}$Zr$_{1/8}$Hf$_{1/8}$Mo$_{1/8}$V$_{1/8}$W$_{1/8}$)C. Such a decrease can be understood by the spectral thermal conductivity $\kappa_{lat}(\omega)$ in **Fig. 5** (a), where phonon modes in the range of 0-3 THz contribute the most to $\kappa_{lat}$ with a significantly reduced tendency, suggesting a higher anharmonic scattering rate of low-frequency phonons. Moreover, the simulated $\kappa_{lat}$ are found to be in good agreement with the experimental measurements (see Fig. 5(b)), demonstrating the accuracy and applicability of our training NEP in studying the thermal conductivities of HECs. In addition, the changes of $\kappa_{lat}$ under temperature were further explored. As displayed in Fig. 5(c), all three HECs possess a decreased $\kappa_{lat}$ with the increase in temperature whereas their tendency ((Ta$_{1/4}$Nb$_{1/4}$Ti$_{1/4}$Zr$_{1/4}$)C > (Ta$_{1/6}$Nb$_{1/6}$Ti$_{1/6}$Zr$_{1/6}$Hf$_{1/6}$Mo$_{1/6}$)C > (Ta$_{1/8}$Nb$_{1/8}$Ti$_{1/8}$Zr$_{1/8}$Hf$_{1/8}$Mo$_{1/8}$V$_{1/8}$W$_{1/8}$)C) remains unchanged at both low and high temperature. Further studies on the mass fluctuations and the strain field fluctuations of the three HECs were conducted to present some clues of decreased $\kappa_{lat}$. As shown in **Table S4**, the mass fluctuations are also monotonically decreased from (Ta$_{1/4}$Nb$_{1/4}$Ti$_{1/4}$Zr$_{1/4}$)C to (Ta$_{1/6}$Nb$_{1/6}$Ti$_{1/6}$Zr$_{1/6}$Hf$_{1/6}$Mo$_{1/6}$)C, then to (Ta$_{1/8}$Nb$_{1/8}$Ti$_{1/8}$Zr$_{1/8}$Hf$_{1/8}$Mo$_{1/8}$V$_{1/8}$W$_{1/8}$)C, which is inconsistent with the previous conclusion of higher mass fluctuations corresponds to higher $\kappa_{lat}$, may implying the minor influence of mass fluctuation on $\kappa_{lat}$ in HECs. In contrast, increased tendencies can be clearly observed in both volumetric and shear strain fluctuations (see Fig. 5(d-g)) at both 300 K and 2300 K, indicating the lowered $\kappa_{lat}$ in HECs may mainly be attributed to the aggravation of strain fluctuations, i.e. lattice distortion.

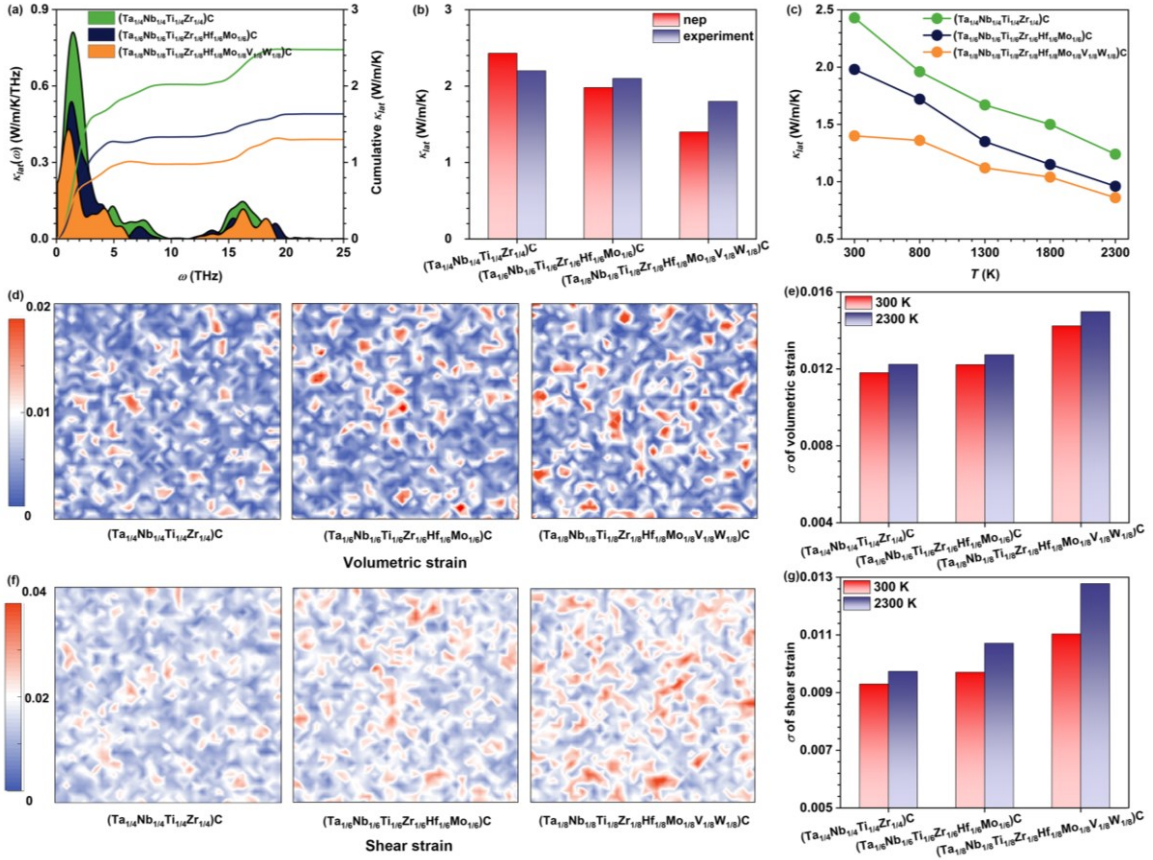

**Figure 5** Thermal conductivities and strain field of $(Ta_{1/4}Nb_{1/4}Ti_{1/4}Zr_{1/4})C$, $(Ta_{1/6}Nb_{1/6}Ti_{1/6}Zr_{1/6}Hf_{1/6}Mo_{1/6})C$, and $(Ta_{1/8}Nb_{1/8}Ti_{1/8}Zr_{1/8}Hf_{1/8}Mo_{1/8}V_{1/8}W_{1/8})C$. (a) $\kappa_{lat}(\omega)$ and the cumulative $\kappa_{lat}$. (b) Comparison of $\kappa_{lat}$ between NEP calculations and our measurements. (c) Predicted $\kappa_{lat}$ as a function of temperature. (d) Volumetric strain distribution on 2D slice of (100) plane. (e) Standard deviations of volumetric strain at 300 K and 2300 K. (f) Shear strain distribution on 2D slice of (100) plane. (g) Standard deviations of shear strain at 300 K and 2300 K.

Besides thermal conductivities, CTEs of HECs were further predicted by MD simulations. **Fig. 6**(a) shows the calculated results and reported experimental values [56-64] of different 1HECs. The theoretical results are in good agreement with experimental reports (within the bounds of experimental measurements), verifying the reliability of our predictions. It should be mentioned that those 1HECs with large predicted CTEs, i.e., MoC, WC, YC, and ScC, are energetically unstable in their FCC phases from a thermodynamic perspective [55]. Moreover, the calculated CTEs of HECs are also comparable to the ones from previous research [33,61] and our measurements with acceptable discrepancies, as listed in **Table 1**, further confirming the feasibility of our MD predictions for HECs. From Fig. 6(b), it is clear to see that CTEs of 4-8HECs are all located within the predicted CTE bound of 1HECs (7.04-20.5 ×$10^{-6}$ /K) with the maximum achieved in $(Ti_{0.25}V_{0.25}Sc_{0.25}Y_{0.25})$C (20.5 ×$10^{-6}$ /K) and the minimum in $(Ti_{0.25}Zr_{0.25}Hf_{0.25}Ta_{0.25})$C (7.52 ×$10^{-6}$ /K). It should be noted that, contrary to the high CTE predicted in the unstable ScC, the similarly high CTE in $(Ti_{0.25}V_{0.25}Sc_{0.25}Y_{0.25})$C is attainable due to the promising stability of $(Ti_{0.25}V_{0.25}Sc_{0.25}Y_{0.25})$C [65], suggesting the feasibility of searching exceptional performance in HECs with our NEP. Additionally, it can also be found from Fig. 6(b) that the rule of mixtures (ROM) method has a tendency to overestimate the CTEs of HECs compared to our MD simulations. From Fig. 6(c), it can be deduced that HECs with fewer TM elements tend to possess larger or smaller CTEs, and the increase in temperature results in an obvious overall increase in CTEs from 300 K to 2300 K.

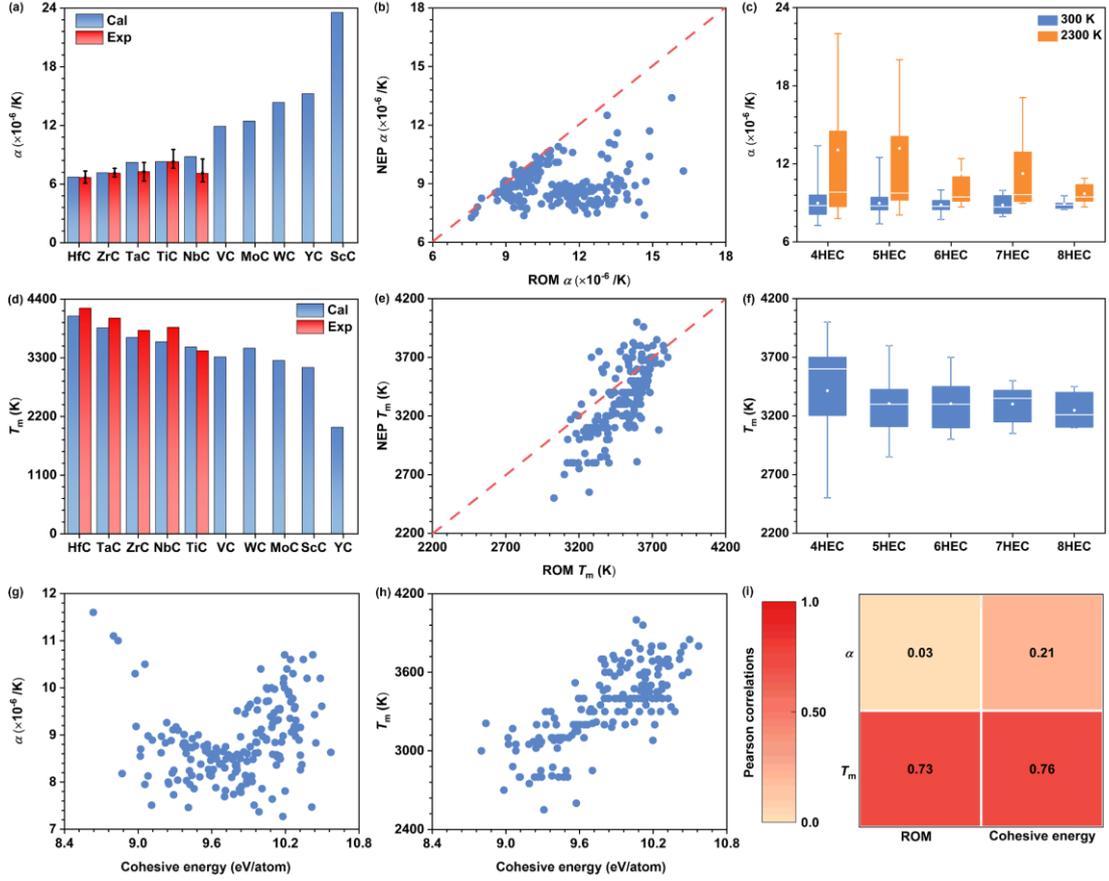

**Figure 6** CTEs and $T_m$ of *n*HECs from NEP calculations, reported experiments, and ROM calculations. (a) CTEs of different 1HECs from NEP calculations and reported experiments [56-64]. Error bars are the maximum and minimum of the reported values. (b) CTEs of different 4-8HECs between NEP and ROM calculations. of *n*HECs from NEP calculations, reported experiments, and ROM calculations. (c) CTEs of 4-8HECs at 300 K and 2300 K. (d) $T_m$ of different 1HECs from NEP calculations and reported experiments [66-69]. (e) $T_m$ of different 4-8HECs between NEP and ROM calculations. (f) $T_m$ of 4-8HECs. (g) Relationship between the cohesive energy and CTEs from NEP calculations. (h) Relationship between the cohesive energy and $T_m$ from NEP calculations. (i) Pearson correlation between different properties.

Table 1 CTEs of HECs from NEP calculations, reported literatures, and our measurements.

| HECs | Exp. ($\times 10^{-6}$/K) | Cal. ($\times 10^{-6}$/K) |
| --- | --- | --- |
| $(Ta_{1/4}Nb_{1/4}Ti_{1/4}Zr_{1/4})C$ | 7.05 | 8.00 |
| $(Ti_{1/5}Zr_{1/5}Hf_{1/5}Ta_{1/5}W_{1/5})C$ | 7.70 (AIMD) [62] | 8.05 |
| $(Ti_{1/5}Zr_{1/5}Hf_{1/5}Nb_{1/5}Ta_{1/5})C$ | 7.85 (MD) [33] | 7.82 |
| $(Ta_{1/6}Nb_{1/6}Ti_{1/6}Zr_{1/6}Hf_{1/6}Mo_{1/6})C$ | 7.07 | 8.25 |
| $(Ta_{1/8}Nb_{1/8}Ti_{1/8}Zr_{1/8}Hf_{1/8}Mo_{1/8}V_{1/8}W_{1/8})C$ | 7.06 | 8.84 |

Since the melting points ($T_m$) of carbides are generally over 3200 K [66-69], it is difficult to measure them directly in experiments. By utilizing MD computations with our trained NEP, the $T_m$ of various HECs can be rapidly predicted. As shown in Fig. 6(d) and **Table S5**, the theoretical predictions for $T_m$ of 1HEC are all in good agreement with reported experiments [66-69] (the average difference is less than 4.72%), indicating the reliability of our trained NEP. From Fig. 6(e), it can be observed that the predicted $T_m$ of HECs are all within the bound of predicted $T_m$ of carbides, where $(Ti_{0.25}Hf_{0.25}V_{0.25}W_{0.25})C$ has the largest $T_m$ (4000 K) among all the predicted 4-8HECs and $(Zr_{0.25}Hf_{0.25}V_{0.25}Y_{0.25})C$ possesses the lowest $T_m$ of 2550 K. Although the $T_m$ prediction of ROM in Fig.6(e) is improved compared to that in CTEs, it still tends to overestimate the values for 4-8HECs. Meanwhile, the $T_m$ distribution of 4HECs is the widest (see Fig. 6(f)), indicating the possibility of achieving higher $T_m$ in HECs with fewer TM elements. Additionally, as previous studies have reported a potential correlation between the cohesive energy and CTEs or $T_\mathrm{m}$ [70], the relationships between them in HECs were also explored. As shown in Fig. 6(g,h), it can be seen that CTEs have no evident correlation with the cohesive energy, while $T_\mathrm{m}$ is predicted to be strongly correlated to the cohesive energy. These results can also be obtained quantitatively by the Pearson correlation coefficients calculated in Fig. 6(i), where almost no correlations can be found for CTEs with ROM and cohesive energy while the cohesive energy-$T_\mathrm{m}$ and ROM-$T_\mathrm{m}$ pairs possess large values, unrevealing the importance of binding and elements on $T_\mathrm{m}$ of HECs.

## 4. Conclusion

In this work, we explore the mechanical and thermal properties of HECs by constructing a general NEP with wide compositional applicability for HECs. To be specific, the efficiency, accuracy, and generalization of the trained NEP for HECs based on the small 1+2HEC training dataset with an equal number of ergodic chemical compositions have been identified, resulting in the construction of a highly accurate and transferable NEP for HECs with low RSMEs of 15.7 meV/atom, 301 meV/Å, and 49.1 meV/atom for energy, force, and virial, respectively. On the basis of MD simulations with our established NEP, mechanical properties and thermal properties, including elastic properties, tensile behaviors, $\kappa_{lat}$, CTEs, and $T_\mathrm{m}$, of HECs are predicted to be consistent with results from first-principle calculations and experimental measurements, further validating the reliability of exploring mechanical and thermal properties of HECs with the well-trained NEP. Our work provides a simple and efficient approach to developing high-entropy ceramics with desirable mechanical and thermal properties.

## Acknowledgments

We acknowledge the financial support from the National Key Research and Development Program of China (No. 2021YFA0715801), the National Natural Science Foundation of China (No. 52122204 and 52402075), and Guangzhou Basic and Applied Basic Research Foundation (SL2023A04J00690).

## Author Contributions

Y. Chu conceived and designed this work. Y. Liu, H. Meng, and H. Yu performed calculations. Z. Zhu and L. Zhuang performed experiments. Y. Chu, H. Yu, and Y. Liu analyzed the data and wrote the manuscript. All authors commented on the manuscript.

## Competing Interests

The authors declare no competing financial interest.

## Data Availability

The data that support the findings of this study are available from the corresponding author upon reasonable request.